\documentclass[11pt]{amsart}
\usepackage{epsfig}

\newcommand{\ul}{\underline}
\newcommand{\charZ}{X({\mathbb Z}\oplus{\mathbb Z})}

\newtheorem{theorem}{Theorem}[section]
\newtheorem{lemma}[theorem]{Lemma}
\newtheorem{proposition}[theorem]{Proposition}

\theoremstyle{definition}

\theoremstyle{remark}

\numberwithin{equation}{section}

\title[Quantization of flat $SU(2)$-connections on torus]{The  Weyl 
quantization and the quantum group quantization of the moduli 
space of flat $SU(2)$-connections on the torus are the same}
\author{R{\u{a}}zvan Gelca}
\address{Department of Mathematics and Statistics, 
Texas Tech University, Lubbock, TX 79409 and Institute of Mathematics
of the Romanian Academy, Bucharest, Romania}
\email{rgelca@math.ttu.edu}
\author{Alejandro Uribe}
\address{Department of Mathematics, 
University of Michigan, Ann Arbor, MI 48109}
\email{uribe@math.lsa.umich.edu}

\date{January 16, 2001}


\begin{document}
\maketitle

\begin{abstract}
We prove that, for the moduli space of flat 
$SU(2)$-connections on the 2-dimensional torus, the
Weyl quantization and the quantization using the 
quantum group of $SL(2,{\mathbb C})$ are the same.
This is done by comparing the matrices of the operators 
associated by the two quantizations to cosine functions. 
We also discuss the 
$*$-product of the Weyl quantization and show that it
satisfies the product-to-sum formula for noncommutative cosines on
the noncommutative torus. 
\end{abstract}

\tableofcontents

\section{Introduction}

Quantization is  a procedure for replacing functions on
the phase space of a physical system (classical observables)
by linear operators. While understood in many general
situations, this procedure is far from being algorithmic. 
Some  more exotic spaces whose quantizations are of 
interest to mathematicians are the moduli spaces 
of flat connections on a surface. 
Among them the case of the moduli space of flat $SU(2)$-connections
on a torus is a particularly simple example of an algebraic 
variety that fails to be a manifold, yet is very close 
to being one. 

In this paper we compare two methods of quantizing the 
moduli space of flat $SU(2)$-connections on the torus. 
The first is the quantization using the quantum group of
$SL(2,{\mathbb C})$. This quantization scheme arose 
when Reshetikhin and Turaev  constructed a 
topological quantum field theory that explains the Jones
polynomial of a knot. It describes both the quantum observables
and the Hilbert spaces in terms of knots and links colored
by representations of the quantum group of $SL(2,{\mathbb C})$.
Heuristically, the operators of the quantization were defined
by Witten using path integrals for the Chern-Simons action.

On the other hand, the moduli space of flat $SU(2)$-connections
on the torus is the same as the character variety of 
$SU(2)$-representations of its fundamental group, so it
admits a covering by the plane. Therefore we can apply
a classical quantization procedure of the plane in an equivariant
manner to obtain a quantization of the moduli space. 
The first such procedure was introduced by Hermann Weyl.
It assigns to each smooth function (classical observable) on the plane
the pseudo-differential operator with symbol equal to the function.

Our main result  is the following

\noindent {\bf Theorem.}{\em 
The Weyl quantization and the quantum group quantization of
the moduli space of flat $SU(2)$-connections on the torus are unitarily
equivalent.  }

The paper is structured as follows. In Section 2 we describe the
geometric realization 
and the K\"{a}hler structure of the
moduli space of flat $SU(2)$-connections on the torus.
In Section 3 we review Witten's description of the quantization
for the particular case of the torus  with a path integral 
of the Chern-Simons action, then explain the rigorous construction
of Reshetikhin and Turaev using quantum groups. Here we also  mention
a result of the first author that  describes the matrices
of the operators associated to cosine functions
in a distinguished basis of the Hilbert space. This distinguished
basis consists of the colorings of the core of the solid torus
by irreducible representations.

We then explain in detail the Weyl quantization of the moduli
space (Section \ref{patru}).
 This is done in the holomorphic setting, which can be related
to the classical, real setting through the Bargman transform.
A distinguished basis of the Hilbert space is introduced  in terms   
of odd theta functions. Section 5 contains the main result of the 
paper (Theorem 5.3).
It shows that the two quantizations are unitarily equivalent.
The unitary equivalence maps the basis consisting of odd theta
functions to the basis consisting of the colored cores of the solid torus.
In Section 6 we discuss the $*$-product that arises from this quantization.
We conclude with some final remarks about the quantization 
scheme based on the Kauffman bracket skein module, for which the 
result does not hold due to a sign obstruction. 

\section{The phase space we are quantizing}

Throughout the paper ${\mathbb T}^2$ will denote the 2-dimensional
torus. 
The  moduli space of flat $SU(2)$-connections on a surface  
is the same as the character variety of $SU(2)$-representations
of the fundamental group of the surface \cite{AB}, i.e. the set of the 
morphisms of the fundamental group of the surface into $SU(2)$ modulo
conjugation.
 This is a complex algebraic variety. In the case of the torus, morphisms
from $\pi_1({\mathbb T^2})={\mathbb Z}\oplus {\mathbb Z}$ to $SU(2)$ 
are parameterized by the images of the two generators of
${\mathbb Z}\oplus {\mathbb Z}$, i.e. by two commuting matrices.
The two  matrices  can be simultaneously 
diagonalized. Moreover conjugation can permute simultaneously  
the  entries in the two diagonal matrices. 
Therefore the character variety is 
\begin{eqnarray*}
\charZ = \{ (s,t)\quad | \quad |s|=|t|=1 \}/(s,t)\sim (\bar{s},\bar{t}).
\end{eqnarray*}
This set is called the ``pillow case''. It  has a 
2-1 covering by the torus, with branching points
$(1,1)$, $(1,-1)$, $(-1,1)$, $(-1,-1)$. We will 
think of the character variety as the quotient of the complex plane  by
the group $\Lambda$ generated by the  translations $z\rightarrow z+1$ and
$z\rightarrow z+i$, and by    the symmetry with respect to
the origin, which we denote by $\sigma$.

 Off the four singularities $\charZ$ is 
a K\"{a}hler manifold, with K\"{a}hler form $\omega$
induced by    ${i\pi}dz\wedge d\bar{z}$ on ${\mathbb C}$. Note that
\begin{eqnarray*}
{i\pi}dz\wedge d\bar{z}=-2i\partial\bar{\partial}\ln h(z,\bar{z}),
\end{eqnarray*}
 where
$h(z,\bar{z})=
e^{-\frac{\pi}{2}|z|^2}$ is
 the weight of the Bargman measure
on the plane. Also, note that the K\"{a}hler form  $\omega$  is the genus 
one case of Goldman's symplectic form defined in  \cite{Go}.

The classical observables are the $C^{\infty}$ functions on this
variety. Using  the covering map we identify the algebra 
of observables on the character
variety  with the algebra of functions on ${\mathbb C}={\mathbb R}
\oplus {\mathbb R}$   generated by
$\cos 2\pi (px+qy)$, $p,q\in {\mathbb Z}$. 
%There is one small technicality
%here, namely we prefer
% to work with the functions $2\cos 2\pi(px+qy)$. There are
%several reasons for putting the factor two in front of the cosine:
% the product-to-sum formula becomes homogeneous,
% the Chebyshev polynomials are defined by $T_n(2\cos x)=2\cos nx$, and
%the equality $2\cos x=e^{ix}+e^{-ix}$ helps avoid multiplications by
%$\frac{1}{2}$ in some of our computations. More important, the 
%functions $2\cos 2\pi(px+qy)$ have a topological meaning in knot theory,
%as we will see below. 

Another family  of important functions on the character variety 
are 
 $\sin 2\pi n(px+qy)/\sin 2\pi (px+qy)$ where $p$ and $q$ are coprime.
As functions on the moduli space, these associate to a connection
the trace  in the $n$-dimensional irreducible representation
of $SU(2)$ of the holonomy of the connection around the curve of slope
$p/q$ on the torus. 

To quantize $\charZ$ means to replace  classical observables $f$ by
linear operators $op(f)$ on some  Hilbert space, satisfying Dirac's conditions:
\begin{itemize}
\item[1.] $op(1)=Id$,
\item[2.] $op(\{f,g\})=\frac{1}{i\hbar}[op(f),op(g)]+O(\hbar)$.
\end{itemize}
Here $\{f,g\}$ is the Poisson bracket induced on $\charZ$ by the 
form $\omega$, which is nothing but
\begin{eqnarray*}
\{f,g\}=\frac{1}{i\pi}\left(\frac{\partial f}{\partial x}\frac{\partial g}{
\partial y}-\frac{\partial f}{\partial y}\frac{\partial g}{
\partial x}\right).
\end{eqnarray*}
Also 
 $[A,B]$ is the commutator of operators, and
$\hbar$ is  Planck's constant. 

Since in our case the phase space
is an orbifold covered by the plane, and this orbifold 
 is K\"{a}hler off singularities,
it is natural to perform   equivariant  quantization of the plane. 
We do this using Weyl's method and then  compare the result 
with the quantization  from \cite{RT}
which was done using the quantum group of $SL(2,{\mathbb C})$.

\section{Review of the  quantum group  approach}

The quantization of the moduli space of flat $SU(2)$-connections
on a surface performed
using the quantum group of $SL(2,{\mathbb C})$ at roots of unity
is an offspring of Reshetikhin and Turaev's construction
of quantum invariants of 3-manifolds \cite{RT}. Their work was inspired by
Witten's heuristic explanation of the Jones polynomial using
Chern-Simons topological quantum field theory. We present below 
Witten's idea for the particular case of the torus, 
and then show how the Reshetikhin-Turaev construction
yields  a quantization of $\charZ$.  

\subsection{Path integrals} 
In \cite{Wi}, Witten outlined a way of quantizing the moduli
space of flat connections on a trivial principal bundle with 
gauge group a simply connected compact Lie group. Let us
recall how this is done when the group is $SU(2)$ and the
principal bundle lies over  the cylinder
over the torus $M={\mathbb T}^2\times [0,1]$. 

For $A$  an
 $SU(2)$-connection on $M$ define the Chern-Simons Lagrangian to be
\begin{eqnarray*}
{\mathcal L}=\frac{1}{4\pi}\int_M \mbox{tr}
(A\wedge dA+\frac{2}{3}A\wedge A\wedge A)
\end{eqnarray*}
where  tr is the trace on the 2-dimensional irreducible representation of
$su(2)$. The Lagrangian is invariant under gauge transformations up
to the addition of an integer. 

The fields we quantize are the flat connections
on ${\mathbb T}^2$, and symplectic reduction restricts our attention
to the moduli space of flat connections \cite{AB}, hence to the character 
variety. Now, using Witten's idea we will associate operators to 
 the 
 observables of the form $2\cos 2\pi(px+qy)$
and $\sin 2\pi n(px+qy)/\sin 2\pi (px+qy)$  
on the torus $p,q,n \in {\mathbb Z}$. 
%Assume first that $p$ and $q$ are coprime.   
%Then, as a function of a flat connection, $2\cos 2\pi(px+qy)$
% is the character of
%the 2-dimensional irreducible representation of $SU(2)$, 
%applied to the holonomy of the connection
%around the curve of slope $p/q$ on the torus. 
%If $p$ and $q$ have a common divisor $n$, say $p=np'$ and $q=nq'$, then
%$2\cos 2\pi (px+qy)$ is the virtual character $\lambda _n=\chi_{n+1}-
%\chi_{n-1}$ 
%applied to the holonomy of the connection around the curve 
%of slope $p'/q'$ on the torus.

%Witten quantized the functions coming from the characters of 
%irreducible representations $SU(2)$. In our notation these
%functions are $\cos 2\pi n(p'x+q'y)/\cos 2\pi (p'x+q'y)$. 
%Our choice for quantizing cosines instead is motivated by the main result
%in \cite{FG}, which shows that the multiplication of quantum observables
%is more elegant if written in terms of noncommutative cosines.
 
Let $N$ be some fixed integer called the level of the quantization. 
As such, Planck's constant is $\hbar =\frac{1}{N}$. 

Assume that $p$ and $q$ are arbitrary integers, and let $n$ be their greatest
common divisor. Denote $p'=p/n$, $q'=q/n$.   
Consider the cylinder over the torus
${\mathbb T}^2\times [0,1]$ and let $C$ be the curve of slope 
$p'/q'$ in ${\mathbb T}^2\times\{ \frac{1}{2}\}$. 
Then the operator associated by the quantization to the function
$\sin 2\pi n(p'x+q'y)/\sin 2\pi (p'x+q'y)$ 
and denoted shortly by $S(p,q)$ ($S$ from sine) is determined 
by the following path integral
\begin{eqnarray*}
<S(p,q) A_1,A_2>=\int_{{\mathcal M}_{A_1,A_2}} e^{iN{\mathcal L}
(A)}\mbox{tr}_{V^n}(\mbox{hol}_C(A)) {\mathcal D}A
\end{eqnarray*}
where $A_1$, $A_2$ are conjugacy classes of flat 
connections on ${\mathbb T}^2$, $A$ is a connection
on  ${\mathbb T}^2\times [0,1]$ such that $A|_{{\mathbb T}^2\times\{0\}}=A_1$
and $A|_{{\mathbb T}^2\times \{1\}}=A_2$, and $\mbox{tr}_{V^n}
(\mbox{hol}_C(A))$, known as the Wilson line, is 
the trace of the $n$-dimensional irreducible representation
of $SU(2)$ evaluated on the  holonomy
of $A$ around $C$ of slope $p'/q'$. Here the ``average'' is taken over all 
conjugacy classes of  connections modulo the gauge group. 

With the same notations 
 one defines the operator $C(p,q)$ ($C$ from cosine)
representing the quantization
of the function $2\cos 2\pi (px+qy) $ by
\begin{eqnarray*}
<C(p,q) A_1,A_2>=\int_{{\mathcal M}_{A_1,A_2}} e^{iN{\mathcal L}
(A)}(\mbox{tr}_{V^{n+1}}-\mbox{tr}_{V^{n-1}}) (\mbox{hol}_C(A)) {\mathcal D}A.
\end{eqnarray*}

We only discuss briefly the Hilbert space of the quantization assuming
the reader is familiar with \cite{atiyah} and \cite{Wi}. Next section will
make these ideas precise. 
The Hilbert space  is spanned by the quantum invariants (i.e. partition
 functions) of all 3-manifolds  with boundary equal to the torus.
Since any 3-manifold can be obtained by performing surgery on a link 
that lies in the solid torus, it follows that the Hilbert space of the 
quantization is spanned by the partition functions of pairs of the form
$(S^1\times {\mathbb D}^2, L)$, where $L$ is a (colored) link
in the solid torus $S^1\times {\mathbb D}^2$. 

\subsection{Quantization using the quantum group of $SL(2,{\mathbb C})$}

The quantization of the 
character variety of the torus using the quantum group of $SL(2,{\mathbb C})$
is a particular consequence  of the topological quantum field theory
constructed in \cite{RT}. Let us briefly describe it.

Fix a level $r\geq 3$ of the quantization, and let $t=e^{\frac{\pi i }{2r}}$.
Comparing with previous section, $N=2r$. 
Quantized integers are defined by the formula 
$[n]=(t^{2n}-t^{-2n})/(t^2-t^{-2})$.  The quantum algebra of 
$sl(2,{\mathbb C})$, denoted by ${\mathbb U}_t$ is a deformation of 
its universal enveloping algebra and has generators $X,Y,K$
subject to the relations
\begin{eqnarray*}
& &   KX=t^2XK, \quad KY=t^{-2}YK,\quad XY-YX=\frac{K^2-K^{-2}}{t^2-t^{-2}},\\
& &   X^r=Y^r=0, \quad K^{4r}=1.
\end{eqnarray*}
This algebra is Hopf, so its representations form a ring under the operations
of direct sum and tensor product. Reducing modulo summands of quantum
trace zero, this ring contains a subring generated by finitely
many irreducible representations  $V^1,V^2, \ldots , V^{r-1}$. Here
$V^k$ has dimension $k$, basis $e_{-(k-1)/2},
e_{-(k-3)/2}\ldots , e_{(k-1)/2}$ and the action of
${\mathbb U}_t$ is defined by 
\begin{eqnarray*}
& & Xe_j=[m+j+1]e_{j+1}\\
 & & Ye_j=[m-j+1]e_{j-1}\\
& & Ke_j=t^{2j}e_j.
\end{eqnarray*}

The idea originating in \cite{KR} and further developed in \cite{RT} is
to color any link in a 3-dimensional manifold by such irreducible
representations. For a knot $K$ we denote by $V^n(K)$ its coloring by
$V^n$. There is a  rule, for which we refer the reader to
 \cite{KR} and \cite{RT},
 for associating numerical
invariants to colored links in the 3-sphere.
Briefly, the idea is to use a link diagram such as the one  in 
Fig. 1, with the local maxima, minima and crossings 
separated by horizontal lines, and then  
define   an automorphism of ${\mathbb C}$ by associating to minima
maps of the form ${\mathbb C}\rightarrow V^n\otimes V^n$, to maxima
maps of the form $ V^n\otimes V^n\rightarrow {\mathbb C}$, to crossings
 the quasitriangular $R$ matrix of the quantum algebra
in the $n$-dimensional irreducible representation, and to parallel
strands tensor products of representations. The automorphism
is then the multiplication by a constant and the link invariant is equal
to that constant. 
For $K$ a knot in the 3-sphere, $V^2(K)$ is its Jones polynomial
\cite{Jones} evaluated at the specific root of unity, and for $n\geq 2$,
 $V^n(K)$ is called  the colored (or generalized) Jones polynomial. 

\begin{figure}[htbp]
\centering
\leavevmode
\epsfxsize=2.1in
\epsfysize=1.8in
\epsfbox{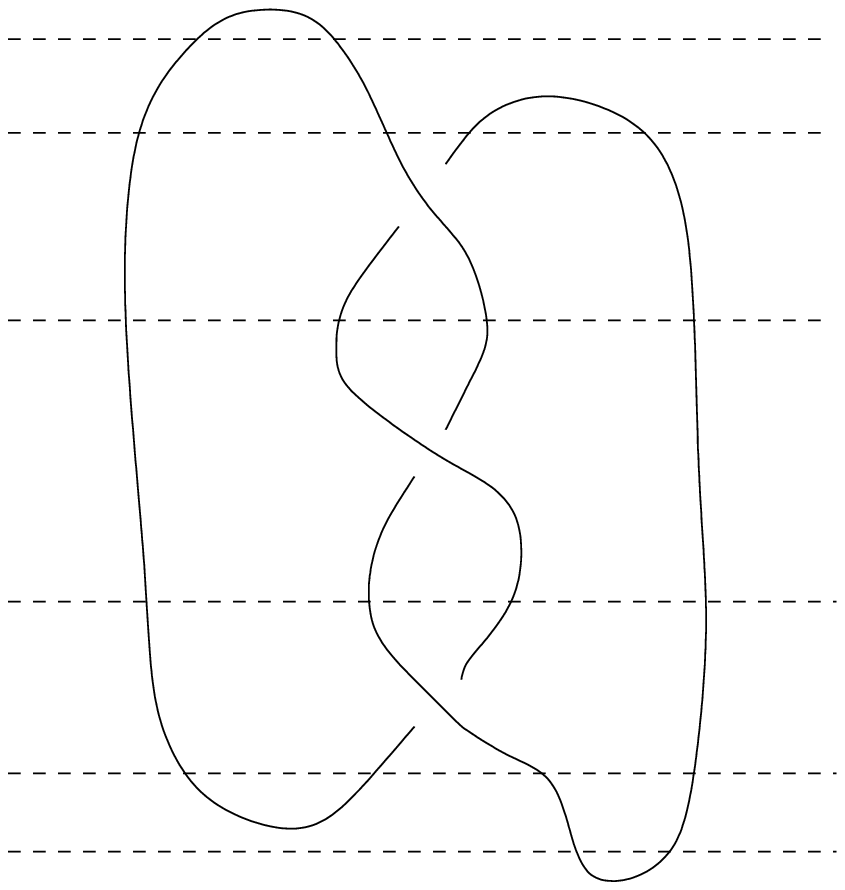}

Figure 1.  
\end{figure}

We are now able to describe  the quantization of
the torus. First consider the vector space freely spanned by all
colored links in the solid torus. On this vector space consider the
pairing $[\cdot, \cdot]$ induced by the operation of
gluing two solid tori such that the meridian of the first is
identified with the longitude of the second and vice versa,
as to obtain a 3-sphere (Fig. 2). 
The pairing of two links $[L_1,L_2]$ is equal to the quantum invariant
of the resulting link in the 3-sphere. The Hilbert space of the quantization
is obtained by factoring the vector space by all linear
combinations of colored links $\lambda$ such that $[\lambda , \lambda']=0$
for any $\lambda'$ in the vector space. This quotient, denoted 
 by $V({\mathbb T}^2)$ by quantum topologists, is finite dimensional.

\begin{figure}[htbp]
\centering
\leavevmode
\epsfxsize=3in
\epsfysize=1.1in
\epsfbox{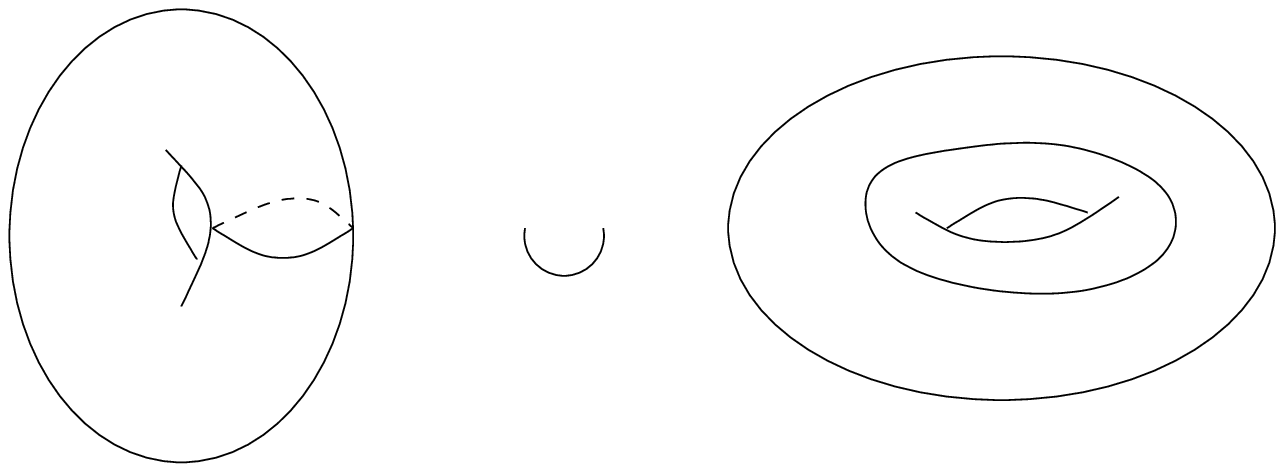}

Figure 2.  
\end{figure}

 Let $\alpha $ be the  core $S^1\times \{0\}$ of 
 the solid torus $S^1\times {\mathbb D}$, ${\mathbb D}=
\{ z, |z|\leq 1\}$. A basis of    $V({\mathbb T}^2)$ is given by 
 $V^{k}(\alpha)$, $k=1,2,\ldots, r-1$. The inner product is determined
by requiring that this basis is orthonormal. The pairing $[\cdot,\cdot]$
is not the inner product. 

Let us point out that these basis elements play an important
role in the construction of the Reshetikhin-Turaev invariants of
closed 3-manifolds. Indeed, as suggested in \cite{Wi} and rigorously
done in \cite{RT}, the quantum invariant of a 3-manifold obtained
by performing surgery on a link is calculated by gluing to the complement
of the link solid tori with cori colored by irreducible representations,
computing the colored link invariants and then summing over all possible
colorings. 

Consider now two integers $p$ and $q$, let $n$ be their greatest 
common divisor,
and let also  $p'=p/n$, $q'=q/n$. The operator $S(p,q)$ associated to the 
classical observable $\sin 2\pi n(p'x+q'y)/\sin 2\pi (p'x+q'y)$ is 
obtained  by coloring the curve of slope $p'/q'$ in the cylinder over the torus
by the representation $V^n$. Also, as explained
 in the previous section, the operator that quantizes
$2\cos 2\pi(px+qy)$ is 
\begin{eqnarray*}
C(p,q)=C(np',nq')=S((n+1)p',(n+1)nq')-S((n-1)p',(n-1)q').
\end{eqnarray*}
Their  action on the Hilbert space of the solid torus 
is defined by  gluing  the cylinder over the torus to the solid 
torus. The operator associated to an arbitrary function
in $C^{\infty}(\charZ)$ is defined by approximating the
function with trigonometric polynomials in cosines, quantizing those,
then passing to the limit.

The matrices of the action can be computed using the pairing 
$[\cdot, \cdot ]$, which now is nondegenerate on $V({\mathbb T}^2)$.
Motivated by the representation theory of  ${\mathbb U}_t$ we extend
formally the definition of the $V^n(K)$ to all integers $n$ by the rules
$V^r(K)=0$, 
$V^{n+2r}(K)=V^n(K)$ and $V^{r+n}(K)=-V^{r-n}(K)$. 
Here the negative sign means that we color the knot by $V^{r-n}$,  then
consider the vector with opposite sign in $V({\mathbb T}^2)$. 
 The following result was proved in \cite{gelca} using the pairing
$[\cdot,\cdot ]$ and
topological quantum field theory with corners.

\begin{theorem}
In any level $r$ and for any integers $p,q$ and $k$ the following formula
holds
\begin{eqnarray*}
C(p,q)V^k(\alpha)=t^{-pq}\left(t^{2qk}V^{k-p}(\alpha)+t^{-2qk}V^{k+p}(\alpha)
\right).
\end{eqnarray*}
\end{theorem}

We will show that the operators of the Weyl quantization act in an
identical way on a basis consisting of theta functions.

\section{The Weyl   quantization}\label{patru}

The first general quantization scheme was introduced by
Weyl in 1931. This scheme applies to functions on ${\mathbb R}^{2n}$
and postulates that the function $e^{2\pi i(px_j+qy_j)}$ corresponds
to the operator $e^{2\pi (pX_j+qD_j)}$ where $X_j$ is multiplication by the
variable $x_j$ and $D_j=\frac{1}{2\pi i}\frac{\partial }{\partial x_j}$.
In general, the operator associated to a function is the pseudo-differential
operator with symbol equal to the function.  Since we are quantizing
a K\"{a}hler manifold, we will  convert to the complex picture using the
Bargman transform.  
 
Here and throughout the paper we choose for Planck's constant
 $\hbar=\frac{1}{N}$, where $N=2r$ is an even  
integer. This is done so that  Weil's integrality condition is satisfied,
and so that the Reshetikhin-Turaev topological quantum
field theory  is well defined.
 The Hilbert space of the quantization is 
the space of square integrable holomorphic sections of a line
bundle with curvature $N\omega$. It is finite dimensional,
as implied by the Heisenberg uncertainty principle since the
phase space is compact. 
It suffices to find the line bundle ${\mathcal L}$ for $N=1$,
and then let the bundle for an arbitrary $N$ be
${\mathcal L}^{\otimes N}$. 

\subsection{The line bundle}
The line bundle ${\mathcal L}$ is defined by a cocycle
\begin{eqnarray*}
\chi :{\mathbb C}\times \Lambda \rightarrow {\mathbb C}\backslash \{0\}
\end{eqnarray*}
 as the quotient ${\mathbb C}\times {\mathbb C}/\sim $ under the  equivalence
  $(z,a)\sim  (w,b)$ if there is $\lambda \in \Lambda $ such that $
(w,b)=(\lambda z, \chi (z,\lambda )a)$. We use the multiplicative notation
since the group $\Lambda $ is not commutative. 
The cocycle condition is
\begin{eqnarray*}
\chi (z,\lambda )\chi(\lambda z , \mu)=\chi (z, \mu \lambda).
\end{eqnarray*}

The cocycle is holomorphic off the singular points of the
character variety, since the line bundle is. 
The compatibility with the hermitian structure yields
$h(z)=|\chi (z,\lambda )| h(\lambda^{-1}z)$.
To find $\chi$ we first determine $\chi (z,m+in)$.   

Since $h(z+m+in)=\exp(-\frac{\pi}{2}(z(m-in )+\overline{z}(m+in)+m^2+n^2))$,
it follows that
 $$|\chi (z, m+in) 
|=\exp\left({\frac{\pi}{2}(z(m-in)+\overline{z}(m+in)+m^2+n^2)}\right).$$
 From the fact that
$\chi $ is holomorphic, it follows that 
$$\chi (z, m+in)=\exp({\pi (z(m-in)+\frac{1}{2}(m^2+n^2))})\cdot 
\exp (i\alpha(m,n)).$$
The cocycle condition yields 
\begin{eqnarray*}
\exp(i\alpha (m,n)+i\alpha (p,q)-i\pi (mq-np))=\exp (i\alpha (m+p,n+q)).
\end{eqnarray*}
This shows that $\exp(\alpha (m,n)-i\pi mn)$ is a morphism from
${\mathbb Z}\times {\mathbb Z}$ to $S^1$. We obtain 
\begin{eqnarray*}
\chi(z, m+in)=(-1)^{mn}\exp (\pi[z(m-in)+\frac{1}{2}(m^2+n^2)])\times \\
\exp (-2\pi i(
\mu m+\nu  n))
\end{eqnarray*}
for some real numbers $\mu$ and $\nu$. 

Recall that $\sigma$ denotes the symmetry of the complex plane  with respect
to the origin. 
We have $|\chi (z, \sigma)|=h(z)/h(-z)=1$.
Since $\chi $ is holomorphic in $z$ it follows that $\chi (z, \sigma)=
\exp (i \pi \beta)$ for some $\beta $. 
We want to determine $\beta $. 

Use the model of the torus obtained by
identifying opposite sides of a square. The action of ${\sigma}$ 
maps  $\frac{1}{2} $ to $-\frac{1}{2}$, and the two correspond to the same
point
on the character variety.
Therefore
\begin{eqnarray*}
\chi \left(\frac{1}{2}, \sigma\right)=\chi \left(\frac{1}{2},
-1\right).
\end{eqnarray*}
We have  
\begin{eqnarray*}
\chi \left(\frac{1}{2}, -1\right) &= &(-1)^0e^{\pi
\left(\frac{1}{2}(-1-0i)+\frac{1}{2}((-1)^2+0^2)
\right)}e^{-2\pi i (-\mu +0\nu)}\\
& = &
e^{2\pi i\mu}.
\end{eqnarray*} 
Hence
$\chi (z, \sigma )=e^{2\pi i \mu }$, so $\mu =\beta /2$.
The same argument with $\chi(\frac{1}{2}i,\sigma)$ and
$\chi(\frac{1}{2}i,-i)$ shows that $\nu=\beta/2$. Also 
\begin{eqnarray*}
\chi \left(\frac{1}{2}+\frac{1}{2}i,\sigma\right)=
\chi\left(\frac{1}{2}+\frac{1}{2}i,-1-i\right)=e^{\pi i[2 (\mu+\nu)-1]}
\end{eqnarray*}
which implies that modulo 2, $2(\mu+\nu) -1=\beta$. Therefore
$\beta =1$.  

 We conclude that 
\begin{eqnarray*}
& & \chi (z, (m+in))
=(-1)^{mn} \exp (\pi[z(m-in)+\frac{1}{2}(m^2+n^2)])\\
& & \chi (z, \sigma)
=-1.
\end{eqnarray*}

\subsection{The Hilbert space of the quantization}
Recall that  the line bundle ${\mathcal L}$  corresponds to the
case where  the Planck's constant is equal to $1$.
To get the general case with $\hbar =1/N$, we consider  the 
line bundle ${\mathcal L}^{\otimes N}$.
 This  bundle  has the hermitian metric defined by
$h_N(z)=(h(z))^N =\exp (-\frac {N\pi}{2}|z|^2)$, and is given by the cocycle
$\chi_N =(\chi )^N$. 
The Hilbert space of the quantization
 consists of the sections of the line 
bundle over ${\mathbb C}$ that  are holomorphic and whose pull-backs to the
plane are  square integrable
with respect to the Bargman measure  
and  satisfy $f(\lambda z)=\chi_N (z, \lambda)f(z)$, $\lambda \in \Lambda$.
The Hilbert space can thus be identified with that 
of holomorphic square integrable
functions on the plane subject to the conditions 
\begin{eqnarray*}
f(z+m+in)=(-1)^{mnN} \exp (\pi N[z(m-in)+\frac{1}{2}(m^2+n^2)])f(z)
\end{eqnarray*}
and 
\begin{eqnarray*}
f(-z)=-f(z).
\end{eqnarray*}
Here the second condition comes from the action of $\sigma$, hence 
if we drop it, we obtain the Hilbert space of the quantization
of the torus. It can be seen  in \cite{BPU} that 
 the torus has other possible quantization
spaces, which arise by  twisting the line bundle with 
flat bundles. This is not the case with $\charZ$. 

As suggested by \cite{BPU} we  replace the Hilbert  space of the
quantization  with 
a space of theta functions.
Let 
\begin{eqnarray*}
\Theta _N=\{f\quad |\quad f(z+m+in)=e^{N\pi(n^2-2inz)}f(z)\}
\end{eqnarray*}
and let 
\begin{eqnarray*}
{\mathcal H}_N=\{f\in \Theta _N\quad | \quad f(z)=-f(-z)\}
\end{eqnarray*}
both endowed with the 
dot product
\begin{eqnarray*}
<f,g>=\int_{{\mathbb T}^2}f(z)\overline{g(z)}e^{-2N\pi y^2}dxdy.
\end{eqnarray*}

Then the Hilbert space of the quantization of the torus
is isomorphic to $\Theta _N$ via the unitary isomorphism
$f(z)\rightarrow e^{-N\pi z^2/2}f(z)$, and the Hilbert space  
of the quantization of $\charZ$ is ${\mathcal H}_N$. 
To be more accurate, the integral that defines the inner product
on ${\mathcal H}_N$ should be performed over a fundamental domain
of the group $\Lambda$, but this gives the same answer as the one above. 

Recall
that an orthogonal basis of $\Theta_N$ is given by
$\theta _j$, 
$j=0,1,\ldots, 2r-1$, where 
\begin{eqnarray*}
\theta _j(z)=\sum_{n=-\infty}^{\infty}e^{-\pi(Nn^2+2jn)+2\pi iz(j+Nn)}.
\end{eqnarray*}
The formula makes sense for all $j$. We have
$\theta _{j+N}(z)=e^{\pi(N+2j)}\theta _j(z)$ and $\theta_{-j}(-z)=
\theta_{j}(z)$, where the second equality follows by replacing $n$ by $-n$.

 An orthonormal basis of  ${\mathcal H}_N$ is given by 
\begin{eqnarray*}
\zeta _j=
\sqrt[4]{\frac{N}{2}}e^{-\pi j^2/N}(\theta _j-\theta_{-j}), \quad j=1,2,\ldots
,r-1.
\end{eqnarray*}
To see why these vectors are indeed orthonormal note that
\begin{eqnarray*}
& & <\theta _k-\theta _{-k},\theta_j-\theta_{-j}>
 =<\theta_k,\theta_j>+
<\theta_{-k},\theta_{-j}>\\
& & -<\theta_{-k},\theta_j>-<\theta_k,\theta_{-j}>=
\delta_{jk}\parallel \theta _j\parallel ^2.
\end{eqnarray*}
Here we used the fact that the $\theta_j$'s can be shifted to have 
 indices equal to one of the numbers  $0,1,\ldots, 2r-1$ and the latter 
form an orthonormal basis in the Hilbert space associated to the 
torus. 

Using the same formula  we
 extend the definition of $\zeta_k$ for all $k\in{\mathbb Z}$.
Clearly $\zeta _r$ has to be equal to zero, since 
\begin{eqnarray*}
\theta_r(z)-\theta _{-r}(z)=\theta _{r}(z)-e^{-\pi(2r-2r)}\theta _r(z)=0.
\end{eqnarray*}
Also, for $1\leq k\leq r-1$ we have 
\begin{eqnarray*}
& \theta _{r+k}-\theta _{-r-k} & =e^{\pi(2r-2(-r+k))}\theta_{-r+k}-
 e^{-\pi(2r+2(r-k))}\theta_{r-k}\\
 && =-e^{2rk\pi}(\theta_{r-k}-\theta_{-r+k}).
\end{eqnarray*}
Thus normalizing we get that $\zeta_{r+k}=-\zeta_{r-k}$.  
Finally,  since $\theta _{j+2r}$ is a multiple of
$\theta _j$ it follows that $\zeta _{j+2r}=\zeta _j$, for all integers $j$.

\subsection{The operators of the quantization}

Each observable $f:\charZ \rightarrow {\mathbb R}$  yields a
sequence of operators indexed by the level $N=2r$. Whenever there is
no danger of confusion we omit the index $N$. We relate Weyl quantization
to Toeplitz quantization and then work with Toeplitz operators, for which
computations are easier.  

Let us first consider the case of the complex plane. Modulo some adjustments to
suit the notations of this paper,   
Propositions 2.96 and 2.97 in \cite{folland} (see also \cite{Hall})
show that the 
operator associated by the Toeplitz 
 quantization to a function $f$ on ${\mathbb C}$
is equal to the  operator associated by the Weyl quantization to the 
function 
\begin{eqnarray*}
\sigma (z,\bar{z})=\frac{1}{N}
\int_{\mathbb C}e^{-2\pi|z-u|^2N}f(u,\bar{u})dud\bar{u}.
\end{eqnarray*}
Note that 
\begin{eqnarray*}
\sigma (z,\bar{z})=e^{\frac{\Delta}{4N}}f(z,\bar{z})
\end{eqnarray*}
 where 
\begin{eqnarray*}
\Delta f=\frac{1}{2\pi}\left(\frac{\partial ^2f}{\partial x^2}+
\frac{\partial ^2f}{\partial y^2}\right).
\end{eqnarray*}
Here we used the notation $z=x+iy$, $\bar{z}=x-iy$. 
 Hence doing  Weyl quantization with symbol $f$ is the same as doing Toeplitz
 quantization
with  symbol  $e^{-\frac{\Delta}{4N}}f$. 
Thinking equivariantly, we now define the operators of the quantization of
the character variety.
Let 
\begin{eqnarray*}
\Pi_N:L^2(\charZ,{\mathcal L}^{\otimes N})\rightarrow {\mathcal H}_N
\end{eqnarray*}
be the  orthogonal projection from the space of square integrable
sections with respect to the measure $e^{-2N\pi y^2}dxdy$
onto  the space  ${\mathcal H}_N$.

To  a $C^\infty$ function $f$ on the character variety  we associate the
operator $op_N(f)$ (in level $N$)  given by 
\begin{eqnarray*}
op_N(f):{\mathcal H}_N\rightarrow {\mathcal H}_N, \quad 
 g\rightarrow \Pi_N\left(\left(e^{-\frac{\Delta}{4N}}f\right)g\right).
\end{eqnarray*}
The operator  $g\rightarrow \Pi_N(f g)$ is the Toeplitz operator
of symbol $f$, denoted by $T_f$. 

An important family of operators are the ones associated to 
the functions $2\cos2\pi(px+qy)$, which we denote by 
$\ul{C}(p,q)$. These are the same as the Toeplitz operators
with symbols 
\begin{eqnarray*}
2e^{-\frac{\Delta}{4N}}\cos 2\pi (px+qy)=2e^{\frac{p^2+q^2}{2N}\pi}
\cos 2\pi (px+qy).
\end{eqnarray*}

\section{Weyl  quantization versus quantum group  quantization}

To simplify the computation, we pull back everything to the line
bundle on the torus. Hence we do the computations
in $\Theta _N$. We start with two lemmas that hold on the
torus. 
They were inspired by \cite{BPU}. 

Let $j,k,p$ be integers such that  $-r+1\leq j,k\leq r-1$,   $p=p_0+\gamma N$
with $\gamma $ an integer. There are two possibilities
$-r+1\leq j+p_0<N=2r$ or $N=2r\leq j+p_0< N+1$. 
Let  also $u(y)$ be a bounded continuous function. 

\begin{lemma}
Assume that $j+p_0<N$. Then
$<e^{2\pi i px}u(y)\theta _j,\theta_k>$ is different from zero if and
only if $k=j+p_0$ and in this case it is equal to 
\begin{eqnarray*}
e^{-\frac{\pi}{2N}p^2+(j+p_0/2)^2/N}
\sum_{m=-\infty}^{\infty}e^{-\pi m^2/2N}
e^{-2\pi i(j+p/2)m/N}
\hat{u}(m)
\end{eqnarray*}
where $\hat{u}(m)$ is the $m$th Fourier coefficient of $u$.  
\end{lemma}

\begin{proof}
Separating the variables we obtain
\begin{eqnarray*}
& & <\exp(2\pi px)u(y)\theta _j,\theta_k>=\int_{{\mathbb T}^2}
\exp(2\pi px)u(y)\theta _j\overline{\theta _k}
e^{-2N\pi y^2}dxdy\\
& & \quad = \sum _{m,n}e^{-\pi(Nm^2+2jm+Nn^2+2kn)}\int_{0}^{1}
e^{2\pi ix(N(m-n)+p+j-k)}dx\\
& & \quad \times \int_{0}^1 e^{-2\pi y(j+Nn+k+Nm)-2\pi Ny^2}u(y)
dy.
\end{eqnarray*}  
The first integral is equal to zero unless $n=m+\gamma$ and $k=p_0+j$. 
If $k=p_0+j$ the expression becomes
\begin{eqnarray*}
& &  e^{-\pi N\gamma ^2-2\pi j\gamma-2\pi p_0\gamma}\int_{0}^{1}
\left(e^{-2\pi (Ny^2+(2j+p_0+N\gamma)y)}\right.\\
& & \quad \times
\left. \sum _m e^{-2\pi \left(Nm^2+Nm\gamma +2m\left(j+\frac{p_0}{2}+Ny
\right)\right)}\right)u(y)dy.
\end{eqnarray*}

After completing
the square in the exponent of the third exponential we obtain that this
is equal to  
\begin{eqnarray*}
e^{-\frac{\pi}{2N}p^2+(j+p_0/2)^2/N}\int_0^1\sum_m
e^{-2\pi N
\left(m+y+\frac{j+p/2}{N}\right)^2}u(y)dy.
\end{eqnarray*}
Using the Poisson formula ($\sum _mf(m)=\sum _m\hat{f}(m)$) for the 
function $e^{-x^2}$  we transform the sum of the exponentials into
\begin{eqnarray*}
\sum_m e^{-\pi m^2/2N}e^{2\pi i\left(y+\frac{j+p/2}{N}\right)m}.
\end{eqnarray*}

It follows that the inner product we are computing is equal to
\begin{eqnarray*}
e^{-\frac{\pi}{2N}p^2+\pi (j+p_0/2)^2/N}
\sum _m e^{-\pi m^2/2N}e^{2\pi i(j+p/2)m/N}
\int_{0}^{1} e^{2\pi i my}u(y) dy
\end{eqnarray*}
which proves the lemma. 
\end{proof}

\begin{lemma}
Assume that $j+p_0\geq N$ and denote  $p_1=p-(\gamma +1)N$. Then 
$<e^{2\pi i px}u(y)\theta _j,\theta_k>$ is different from zero if and
only if $k=j+p_1$ and in this case it is equal to 
\begin{eqnarray*}
e^{-\frac{\pi}{2N}p^2+\pi(j+p_1/2)^2/N}\sum_{n=-\infty}^{\infty}e^{-\pi m^2/2N}
e^{-2\pi i(j+p/2)m/N}
\hat{u}(m)
\end{eqnarray*}
where $\hat{u}(m)$ is the $m$th Fourier coefficient of $u$.  
\end{lemma}

\begin{proof}
We start with a computation like the one in the proof of Lemma 1
to conclude that $k=p_0+j-N=p_1+j$ and $m=n-\gamma -1$.
From here the same considerations apply mutatis mutandis
to yield the conclusion. 
\end{proof}

\begin{theorem}
There exists a unitary isomorphism between $V({\mathbb T}^2)$ and
${\mathcal H}_N$ sending $V^j(\alpha)$ to $\zeta_j$,
 $j=1,2,
\ldots , r-1$, 
which transforms the  operator associated through quantum group 
quantization to  the function  $f$
into the operator associated by Weyl quantization to the same
function. 
\end{theorem}

\begin{proof}
We  verify that the matrix of the operator $\ul{C}(p,q)$ in the basis
$\zeta_j$ is the same as the matrix of the operator
$C(p,q)$ in the basis $V^{j}(\alpha)$. We have
\begin{eqnarray*}
\ul{C}(p,q)=e^{\frac{p^2+q^2}{2N}}T_{2\cos 2\pi(px+qy)}
\end{eqnarray*}
where $T_{2\cos 2\pi (px+qy)}$ is the Toeplitz operator of symbol
$2\cos 2\pi (px+qy)$. Let us   pull back everything 
to the torus using the covering map ${\mathbb T}^2\rightarrow \charZ$ so that
we can work with exponentials.

We do first the case $j+p_0<N$. If in Lemma 5.1  we let $u(y)=e^{2\pi iqy}$
we obtain 
\begin{eqnarray*}
T_{e^{2\pi i px +2\pi i qy}}\theta _j=e^{-\frac{\pi}{2N}p^2+(j+p_1/2)^2/N}
e^{-\pi q^2/2N}e^{-2\pi i (j+p/2)q/N}\theta _{j+p_0}.
\end{eqnarray*}
 Using this formula and the fact that 
\begin{eqnarray*}
\zeta_j=\sqrt[4]{\frac{N}{2}}e^{-\pi j^2/N}(\theta _j-\theta _{-j})
\end{eqnarray*}
after doing the  algebraic computations we arrive at 
\begin{eqnarray*}
e^{\frac{p^2+q^2}{2N}\pi}T_{2\cos 2\pi (px +qy)}\zeta _j=
t^{-pq}\left(t^{2jq}\zeta _{j-p_0}+t^{-2jq}\zeta _{j+p_0}\right).
\end{eqnarray*}
If  $j+p_0\geq N$, we let  $p_1=p-(\gamma +1)N$, and 
 an application of Lemma 5.2  shows that in this case 
\begin{eqnarray*}
e^{\frac{p^2+q^2}{2N}\pi}T_{2\cos 2\pi (px +qy)}\zeta _j=
t^{-pq}\left(t^{2jq}\zeta _{j-p_1}+t^{-2jq}\zeta _{j+p_1}\right).
\end{eqnarray*}
But we have seen that $\zeta _{j+N}=\zeta_j$ for all $j$, so in the 
above formulas $p_0$ and $p_1$ can be replaced by $p$. It
follows that for all $j$, $1\leq j\leq r-1$, and all integers $p$ and
$q$ we have
\begin{eqnarray*}
e^{\frac{p^2+q^2}{2N}\pi}T_{2\cos 2\pi (px +qy)}\zeta _j=
t^{-pq}\left(t^{2jq}\zeta _{j-p}+t^{-2jq}\zeta _{j+p}\right).
\end{eqnarray*}
 
Theorem 3.1 shows   that
\begin{eqnarray*}
C(p,q)V^j(\alpha)
=t^{-pq}\left(t^{2jq}V^{j-p}(\alpha)+t^{-2jq}V^{j+p}(\alpha)
\right).
\end{eqnarray*}
Hence the unitary isomorphism defined by $V^{j}(\alpha)\rightarrow
\zeta _j$ 
transforms the operator $C(p,q)$ into the operator  $\ul{C}(p,q)$, and
the theorem is proved. 
\end{proof}

From now on we identify the two quantizations and use the notation
$C(p,q)$ for the operators. 
Recall the notation $t=e^{i\pi/N}$.
As a byproduct of the proof of the theorem we obtain
the following product-to-sum formula for  $C(p,q)$'s, which was already noticed
in \cite{gelca}. 
\begin{proposition}
For any integers $m,n,p,q$ one has
\begin{eqnarray*}
C(m,n)*C(p,q)=
 t^{|^{mn}_{pq}|}C(m+p,n+q)+
t^{-|^{mn}_{pq}|}C(m-p,n-q),
\end{eqnarray*}
where $|^{mn}_{pq}|$ is the determinant.
\end{proposition}
%Since this formula relies only on the computations 
%from the proof of the theorem, it is valid regardless of
%whether $N$ is even or odd. 
We conclude this section by noting that in Witten's picture 
 the operator associated
by Weyl quantization to the function $\sin2\pi(n+1)(p'x+q'y)/\sin2\pi(p'x+q'y)$
($p',q'$ relatively prime) is the same as the quantum group quantization
of the Wilson line around the curve of slope $p'/q'$ on the torus
in the $n$-dimensional irreducible representation of $SU(2)$.  

%For those familiar with quantum invariants of knots let
%us mention also that 
% as a consequence of the theorem we can  derive an integral formula
%for the $n$th colored Kauffman bracket of a $(p,q)$-torus
%knot (this time $p$ and $q$ have to be coprime):
%\begin{eqnarray*}
%\kappa _n(K) = \sum_{k\leq \lfloor \frac{n}{2}\rfloor}
%<T_{2\cos [2\pi (n-2k)(px+qy)]}\zeta _1, \Omega>
%\end{eqnarray*}
%where in the above sum $T_0$ should be replaced by $1$ (not by $2$), and
%\begin{eqnarray*}
%\Omega =\sum_{j=1}^{r-1}(-1)^{j-1}\frac{t^{2j}-t^{-2j}}{t^2-t^{-2}}\zeta _j.
%\end{eqnarray*}
%Of course, after computing the integrals one arrives at the same formula
%as the one given in \cite{FG}. 

\section{The star product}

\subsection{Definition of the star product}

Let $(M,\omega)$ be a symplectic manifold. A $*$-product on $M$ is a binary
operation on 
\begin{eqnarray*}
C^\infty (M)[[N^{-1}]]
\end{eqnarray*}
which is associative, and for all $f,g\in C^\infty (M)$ satisfies
$N^{-k}f*g=f*N^{-k}g=N^{-k}(f*g)$ and also
\begin{eqnarray*}
f*g=\sum _{k=0}^{\infty}N^{-k}B_k(f,g).
\end{eqnarray*}
The operators $B_k(f,g)$ are bi-differential operators from
 $C^{\infty}(M)\times C^\infty (M)$ to
$C^\infty (M)$, such that 
$B_0(f,g)=fg$,  and such that   Dirac's  correspondence principle 
\begin{eqnarray*}
B_1(f,g)-B_1(g,f)=\{f,g\}
\end{eqnarray*}
is satisfied.
Here $\{f,g\}$ stands for the Poisson bracket induced by the
symplectic form. One  says that $C^{\infty}(M)[[N^{-1}]]$ is a deformation
of $C^\infty (M)$ in the direction of the given Poisson bracket. 
We use $N$ for the variable of the formal series to be consistent 
with the rest of the paper. 

%For a given star product, a linear map $\tau :{\mathcal A}\rightarrow 
%{\mathbb C}[[N^{-1}]]$ is a $*$-compatible trace  if for all 
%$f,g\in C^\infty (M)$, $\tau (N^{-i}f)=N^{-i}\tau (f)$ and 
%$\tau (f*g)=\tau (g*f)$ and
%\begin{eqnarray*}
%\tau{f}=t^n\sum_{i=0}^\infty A_i(f)N^{-i}, \quad \mbox{where} \quad 
%A_0(f)=\int_M\frac{f\omega ^n}{n!}.
%\end{eqnarray*}

The character variety  $\charZ$  is  a
symplectic manifold off the four singularities. 

\begin{proposition} The  formula 
\begin{eqnarray*}
& & 
2\cos2\pi(mx+ny)*2\cos 2\pi (px+qy)\\
\quad & & = t^{|^{mn}_{pq}|}2\cos 2\pi ((m+p)x+
(n+q)y)\\
& & \quad + t^{-|^{mn}_{pq}|}2\cos 2\pi ((m-p)x+
(n-q)y)
\end{eqnarray*}
defines a $*$-product on $ C^\infty{\charZ} (M)
[[N^{-1}]]$, 
which is a deformation quantization in the direction of the 
K\"{a}hler form $i\pi dz\wedge d\bar{z}$. %The functional  $\tau$, with
%$\tau (1)=1$ and $\tau (2\cos 2\pi (mx+ny))=0$ if $m\neq 0$ or $n\neq 0$ is
%a $*$-compatible trace.  
\end{proposition}

In these formulas  the exponentials should
be expanded formally into power series in $N^{-1}$. 

\begin{proof}
We have 
\begin{eqnarray*}
& & 2\cos 2\pi (mx+ny)*2\cos 2\pi (px+qy)\\ & & \quad - 
2\cos 2\pi (mx+ny)*2\cos 2\pi (px+qy)=\\
& &{ \pi (imq-inp)}N^{-1}2\cos 2\pi ((m+p)x+(n+q)y)\\
&  & +{\pi (inp-imq )}N^{-1}2\cos 2\pi((m-p)x+(n-q)y)\\ & & -
{\pi(ipn-iqm)}N^{-1}2\cos 2\pi((p+m)x+(q+n)y)\\ & &  -
{\pi(imq-inp)}{N^{-1}}2\cos 2\pi ((p-m)x+(q-n)y)+O(N^{-2})\\
& & = 2\pi i N^{-1}(mq-np)2\cos 2\pi ((m+p)x+(n+q)y)\\ & & -
2\pi i N^{-1} (mq-np)2\cos 2\pi (( m-p)x+(n-q)y)+ O(N^{-2})\\
& & =  N^{-1}\{2\cos 2\pi (mx+ny),2\cos 2\pi (px+qy)\}+O(N^{-2}).
\end{eqnarray*}
so the correspondence principle is satisfied. The coefficients $B_k(f,g)$
are bidifferential operators since
\begin{eqnarray*}
B_1(f,g)=\frac{1}{4\pi i} \det \left|
\begin{array}{clcr}
\frac{\partial }{\partial x_1} & \frac{\partial}{\partial y_1}\\
\frac{\partial }{\partial x_2} & \frac{\partial}{\partial y_2}
\end{array}
\right|
f(x_1,y_1)g(x_2,y_2)\vert _{\stackrel{x_1=x_2=x}{\scriptstyle{y_1=y_2=y}}}
\end{eqnarray*}
and $B_k=B_1^k/k!$. 
%The compatibility of the trace with the $*$-product can be checked directly,
%but it is a consequence of the facts from next section.
\end{proof}

We would like to point out that this $*$-product is different from the one
that would arise if we applied the quantization methods outlined 
in \cite{berezin} since Berezin's ideas correspond to the 
anti-normal (respectively normal) ordering of the operators. 

Thinking now  of the deformation parameter as a fixed natural number
we see that the $*$-algebra defined in  Proposition 6.1 is a
subalgebra of Rieffel's  noncommutative torus \cite{Ri}
with rational Planck's constant.
That is, our $*$-product is the restriction of Rieffel's $*$-product
to  trigonometric series in cosines.  

\subsection{The generalized Hardy space}

In \cite{Gu} Guillemin has shown that for each compact symplectic prequantizable
manifold $M$ there exists a $*$-product
and a circle bundle $S^1\hookrightarrow P\rightarrow M$ with a
canonical volume $\mu$ such that $L^2(P,\mu)$ contains finite
dimensional vector spaces ${\mathcal H}_N$ of $N$-equivariant
functions (mod the action of $S^1$) satisfying for all
$f,g\in C^{\infty}(M)$:
\begin{eqnarray*}
\Pi_NM_f\Pi_NM_g\Pi_N=\Pi_NM_{(f*g)_N}\Pi_N+O(N ^{-\infty}).
\end{eqnarray*}
where $\Pi_N$ is the orthogonal projection onto ${\mathcal H}_N$,
$M_f$ is the multiplication by $f$ and $(f*g)_N$ is
 to be understood 
as the $*$-product  for a certain fixed integer value of $N$.

Our operators are not Toeplitz so they won't fit exactly Guillemin's 
construction. However, along the same lines we will construct a 
representation of the $*$-algebra from the previous section onto
an infinite dimensional Hilbert space that contains all ${\mathcal H}_N$'s
as direct summands. 
 The reader can find
a detailed account on how these things are done in general in \cite{Bor}.

Let ${\mathcal L}$ be the line bundle constructed in 4.1, and
let $Z\subset {\mathcal L}^*$ be  
 the unit circle bundle in the dual of 
${\mathcal L}$. $Z$ is an $S^1$-principal bundle. 
A point in $Z$ is a pair $(x,\phi)$, where $x\in {\mathbb C}$ and  $\phi$ is a 
complex valued functional with $|\phi(x)|=\| x\|$ (the length of 
$x$ being given by the hermitian structure). More precisely
\begin{eqnarray*}
Z=\{(z,\xi); \quad |\xi |=e^{-\pi |z|^2/2}\}.
\end{eqnarray*}
The map $(z,\xi)\rightarrow (z,\xi/|\xi|)$ identifies $Z$ with 
${\mathbb C}\times S^1$. Let $\theta $ be the argument of $\xi$
and consider the  volume form $d\theta dz$ on $Z$. Using this volume form 
we can define the space $L^2(Z)$. 

Now let $N$ be an even integer. 
 Note that
\begin{eqnarray*}
{\mathcal L}^{\otimes N}\simeq Z\times_N{\mathbb C}
\end{eqnarray*} 
where $Z\times _N{\mathbb C}$ is the quotient of $Z\times {\mathbb C}$ 
by the equivalence
\begin{eqnarray*}
(p\cdot e^{iN\theta},z)\sim (p, e^{iN\theta }z).
\end{eqnarray*}
A complex valued smooth function $f$ on $Z$ is called $N$-equivariant
if for all $(x,\phi)\in Z$, 
\begin{eqnarray*}
f(x,\phi\cdot e^{ i\theta})=e^{iN\theta }f(x,\phi).
\end{eqnarray*}
The set of all $N$-equivariant functions is denoted by
$C^{\infty}(Z)_N$. There exists an isomorphism
\begin{eqnarray*}
 C^\infty({\charZ},{\mathcal L}^{\otimes N})\simeq
C^\infty (Z)_N,
\end{eqnarray*}
which transforms a section $s\in  
C^\infty({\charZ},{\mathcal L}^{\otimes N})$ to a function
$f$, with $f(z,\xi)=\xi ^Ns(z) $. Here of course $s$ has to be viewed
as a ${\mathbb C}$-valued function subject to the equivariance conditions
from Section 4. It is easy to see that this map gives rise
to a unitary isomorphism between the space of $L^2$ sections 
of ${\mathcal L}^{\otimes N}$ and the $L^2$-completion of
$C^\infty (Z)_N$. A little Fourier analysis (involving integrals of
the form $\int e^{im\theta }e^{-in\theta }d\theta $ )  shows that for
different $N$'s, the images of the corresponding $L^2$ spaces
are orthogonal. 

As a result, the spaces ${\mathcal H}_N$ are embedded as 
 mutually orthogonal subspaces of $L^2(Z)$. Define 
\begin{eqnarray*}
{\mathcal H}=\bigoplus _{N \mbox{ even}} {\mathcal H}_N.
\end{eqnarray*}
This space is the  generalized version of the classical
Hardy space.  

We denote by  $\Pi$ the orthogonal projection of $L^2(Z)$ onto ${\mathcal H}$.
Let $f$ be a smooth  function on $\charZ$, which  can be viewed as
as the limit (in the $C^\infty$ topology) of a sequence of trigonometric
polynomials in $\cos (px+qy)$, $p,q\in {\mathbb Z}$.  Define the 
operator 
\begin{eqnarray*}
op(f) :{\mathcal H}\rightarrow {\mathcal H}, \quad 
g\rightarrow  \Pi \left(\left(e^{-\frac{\Delta}{4N}}f\right)g \right).
\end{eqnarray*}
If  $e^{-\frac{\Delta}{4N}}f$ is a $C^\infty$ function  on 
the character variety, this this  operator is bounded.
The restriction of $op(f)$ to ${\mathcal H}_N$ coincides with 
$op_N(f)$, for all $N\geq 1$. 
Moreover, the product-to-sum formula
from  Proposition 5.4 shows that the $*$-product 
  on $C^{\infty }(\charZ)[[N^{-1}]]$ defined by the multiplication of
these operators is the same
as the $*$-product introduced in Section 6.1. 

\section{Final remarks} 

An alternative approach to the Reshetikhin-Turaev theory was constructed
in \cite{BHMV} using Kauffman bracket skein modules. 
This approach also leads to a quantization of the moduli space of flat
$SU(2)$ connections on the torus. Do we obtain the Weyl quantization
in that situation as well? The answer is no. 

Indeed, the analogues of the basis  vectors $V^n(\alpha)$ 
are the colorings of the core of the solid torus by Jones-Wenzl 
idempotents. More precisely, to the  
vector $V^n(\alpha)$ corresponds the vector
$S_{n-1}(\alpha)$, where $S_{n-1}(\alpha)$ is the coloring of 
$\alpha$ by the $n-1$st Jones-Wenzl idempotent. 
It follows from Theorem 5.6 and the discussion preceding it in \cite{FG}
that the action of the operator associated to $2\cos 2\pi (px+qy)$, denoted
by $(p,q)_T$, on these basis elements is
\begin{eqnarray*}
(p,q)_TS_{n-1}(\alpha)=(-1)^qt^{-pq}(t^{2qk}S_{k-p-1}(\alpha)+t^{-2qk}
S_{k+p-1}(\alpha)).
\end{eqnarray*}
The factor $(-1)^q$ does not appear in the formula from Theorem 3.1,
proving  that this quantization is different. However, both quantizations
yield the same $*$-algebra, as shown  in \cite{FG}. 

Finally, although quantum field theory is intimately related to 
Wick quantization, the present  paper shows that this is not
the case with the topological quantum field theory of Reshetikhin and Turaev.

\end{document}